\documentclass[a4paper]{jpconf}
\usepackage{graphicx}
\usepackage{epsfig}
\def\lsim{\mathrel{\rlap{\lower4pt\hbox{\hskip1pt$\sim$}}
    \raise1pt\hbox{$<$}}}                % less than or approx. symbol
\def\gsim{\mathrel{\rlap{\lower4pt\hbox{\hskip1pt$\sim$}}
    \raise1pt\hbox{$>$}}}                % greater than or approx. symbol

\def\sba2{\sin ^2 (\beta - \alpha)}
\def\cba2{\cos ^2 (\beta - \alpha)}

\def\r{\rightarrow}

\def\nue{\nu_{\mathrm{e}}}

\def\h{\mathrm h}
\def\A{\mathrm A}
\def\W{\mathrm W^{\pm}}
\def\Z{\mathrm Z}

\def\H{\mathrm H}
\def\Hpm{\mathrm H^\pm}
\def\WW {\mathrm W^+ \mathrm W^-} 
\def\ZZ {\mathrm Z \mathrm Z} 
\def\ee{\mathrm e^+\mathrm e^-}

\def\nn{\nu \bar{\nu}}
\def\qq{\mathrm q \bar{\mathrm q}}

\def\tautau{\tau^{+}\tau^{-}}

\def\ellell{\ell^{+}\ell^{-}}

\def\bb{\mathrm b \bar{\mathrm b}}

\def\pb{\mathrm{pb}^{-1}}

\def\Gcs{\mathrm{GeV/c}^2}

\def\mH{m_{\H}}

\def\mh{m_{\h}}
\def\mA{m_{\A}}

\def\mHpm{m_{\Hpm}}
\def\mZ{m_{\Z}}

\def\mt{m_{\mathrm t}}

\def\sqs{\sqrt{s}}

\def\mA{m_\mathrm{A}}

\begin{document}
\title{Higgs boson searches at LEP}

\author{P Teixeira-Dias\\On behalf of the LEP Higgs working group}

\address{Department of Physics, RHUL -- Royal Holloway, University of
  London}

\ead{pedro.teixeira-dias@rhul.ac.uk}

\begin{abstract}
In this paper we report on the legacy of Higgs boson searches at
LEP. Specifically, the results of the statistical combination of the
searches carried out by the ALEPH, DELPHI, L3 and OPAL experiments are
presented. In the search for the Standard Model (SM) Higgs boson, a
signal with $\mh<114.4\,\Gcs$ has been excluded at the 95\% confidence
level (CL) or higher.
The LEP collaborations also carried out extensive searches for Higgs
particles predicted by many scenarios beyond the Standard Model. Here
we can only report on a very small fraction of these searches and
refer the reader to the complete list of LEP-combined search results.

\end{abstract}
%
%\mbox{ }\vspace{-1.0cm}
\section{Introduction}
During the decade when it operated, the Large Electron-Positron (LEP)
collider at CERN was at the forefront of the search for evidence of
Higgs bosons, which are predicted by the Higgs mechanism to be at the heart of
the explanation for the masses of elementary particles.
The four LEP experiments carried out extensive searches for the Higgs
scalar predicted in the context of the Standard Model (SM).  More
complex Higgs sectors, such as those predicted by Two Higgs Doublet
Models (THDM) or comprising scalars with unusual decay modes (e.g.,
fermiophobic Higgs, invisible Higgs) were also searched for.
In the second phase of LEP (from 1996 onwards, at centre-of-mass
energies above the W pair production threshold) the sensitivity to a
possible signal was increased by maximizing the integrated luminosity
of the collected data samples, as well as the $\ee$ centre-of-mass
energy $\sqrt{s}$, which was increased every year. In the year 2000,
its final year of operation, LEP2 delivered data to the experiments at
energies up to $\sqs=$209\,GeV. Additional sensitivity to possible
signals was obtained by statistically combining the search results of
the four individual experiments. The combined LEP2 data sample used by
the LEP Higgs working group totalled $2461\,\pb$ collected between
$\sqs=189-209$\,GeV (of which $536\,\pb\,(32.5\,\pb)$ were accumulated
at $\sqs\ge{206\,(208)\,}$GeV).
%(see Table \ref{tab-LEPdata}).
%
%\begin{table}[htbp]
%\caption{\small Integrated luminosities of the data samples of the
%  four experiments and their sum (LEP). The subsets taken at energies
%  exceeding 206\,GeV and 208\,GeV are listed separately.
%\label{tab-LEPdata}}
%\begin{center}
%\begin{tabular}{lccccc}
%\br
%%\multicolumn{6}{c}{Integrated luminosities in $\pb$}\\
%                          & ALEPH      & DELPHI     & L3        & OPAL   & LEP \\
%\mr
%$\sqs\ge 189$\,GeV      & 629        & 608        & 627       & 596    & 2461  \\
%$\sqs\ge 206$\,GeV      & 130        & 138        & 139       & 129    & 536   \\
%$\sqs\ge 208$\,GeV      & 7.5        & 8.8        & 8.3       & 7.9    & 32.5  \\
%\br
%\end{tabular}
%\end{center}
%\end{table}
%
\section{The search for the SM Higgs boson}
At LEP the main process for SM Higgs production is the Higgsstrahlung
process, $\ee\r\H\Z$, which has a kinematic threshold at
$\mH=\sqs-\mZ$. (Small additional contributions to Higgs production,
via WW/ZZ fusion ($\ee\r\nue\bar{\nue}\H/\ee\H$) allow some additional
sensitivity beyond this threshold.) The main search topologies are
therefore dictated by the dominant Higgs decay modes (mostly $\bb$,
some $\tautau$) and the Z decay modes. All four LEP experiments
carried out searches for $(\H\r\bb)(\Z\r\ellell,\,\nn,\,\qq)$
(respectively: the leptonic\footnote{Here we use the symbol $\ell$ and
  the word ``lepton'' to denote an electron or a muon.}, missing
energy, and four-jet topologies), and for the main topologies with taus:
$(\H\r\tautau)(\Z\r\qq)$ and $(\H\r\bb)(\Z\r\tautau)$. Here we report
on the LEP-wide combination\,\cite{lep_smh}, which is based on the
final results of the Higgs search by the individual
collaborations\,\cite{a_smh,d_smh,l_smh,o_smh}.

The search backgrounds were reduced by applying suitable event
selection criteria including, crucially for most search channels, the
ability to experimentally tag b-jets. Typically, after selection the
remaining background consisted mostly of $\ZZ$ and $\WW$ events
(including small contributions from $\Z\ee,\,\Z\nn,\,\W\mathrm{e}\nu$)
and $\qq$ events with additional gluon radiation.

The combination of the results of the four experiments is based on the
ratio of the extended likelihoods for the signal-plus-background
hypothesis and the background-only hypothesis (full details available
in Ref.\,\cite{lep_smh}):
% enter LR here
\begin{equation}
Q=\frac{\cal{L}_{\mathrm{s+b}}}{\cal{L}_{\mathrm{b}}},
\phantom{XXX}-2\ln{Q(\mh)}=2s_{\mathrm{tot}}(\mh)-2\sum_i\ln(1+(\frac{s(\mh)}{b})_i)
\end{equation}
where $s_{\mathrm{tot}}$ is the total expected signal in a given
search channel (i.e., for a given centre-of-mass energy, search
topology, experiment) and the sum is over all the candidate events
selected in the data sample.
Each candidate event $i$ selected in the data contributes a weight
$\ln(1+s/b)$ to the test statistic $-2\ln{Q}$, where $s$ and $b$ are
the signal and background expected in the bin of some discriminating
variable where the candidate lies. (1D and 2D discriminants based on
e.g., reconstructed Higgs mass, b-tagging probability, neural network
output, were used.)
Independent search channels are combined by adding the respective
$-2\ln{Q}$ contributions.
In order to determine robustly how background-like or
signal-plus-background-like an observed result is, it is compared with
the outcome of a large number of toy MC experiments.
Statistical and systematic uncertainties are included
in the combination process, and have been verified to have a small
effect in the results.

\begin{center}
\begin{figure}[h]
\begin{minipage}{18pc}%{14pc}
\hspace{2pc}\includegraphics[width=15.05pc]{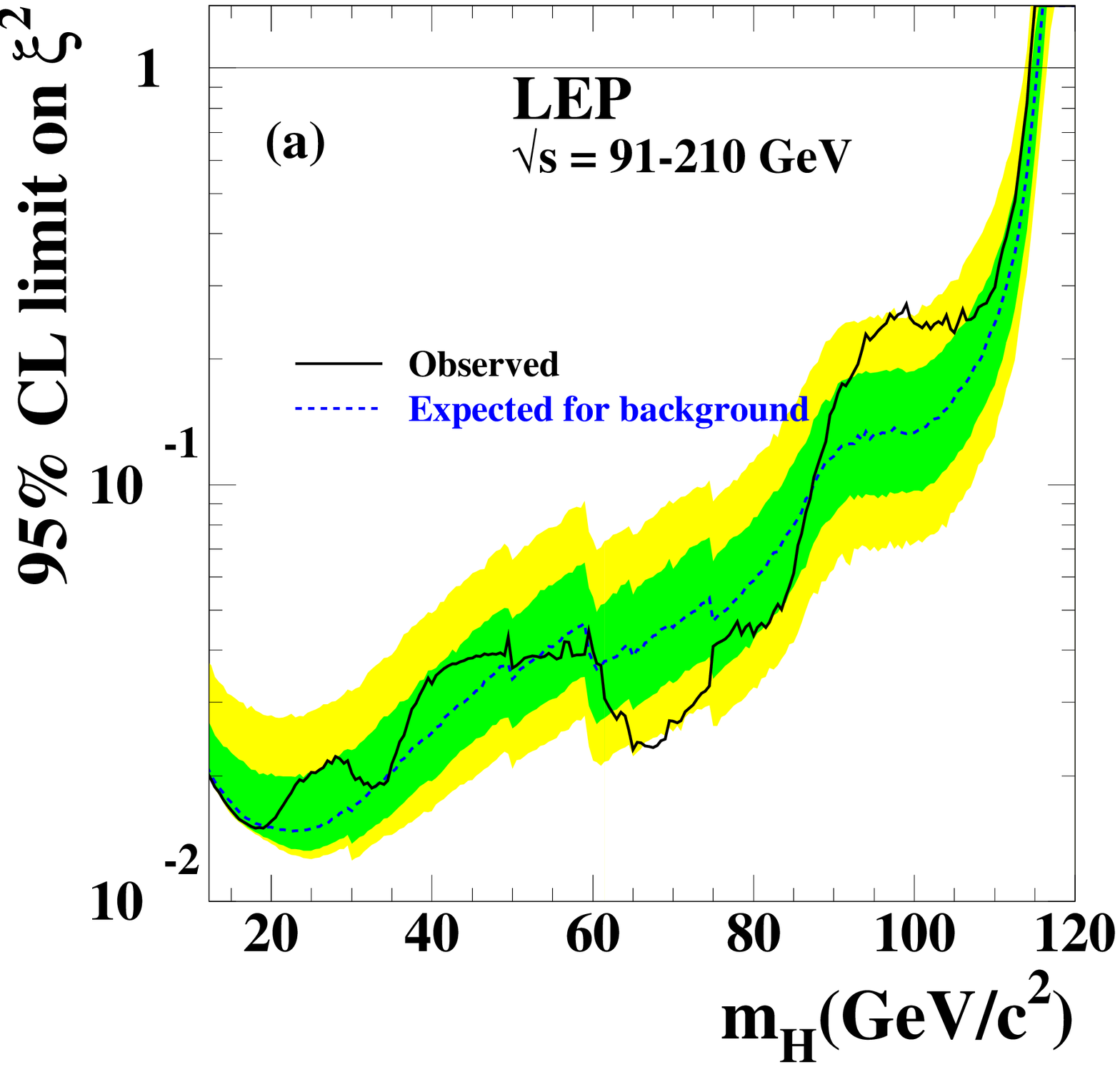} %18.2pc
\caption{\label{fig-one}95\%\,CL limit on the hZZ coupling in
  non-standard
  models. $\xi^2=(g_{\small\mathrm{hZZ}}/g^{\small\mathrm{SM}}_{\small\mathrm{hZZ}})^2$,
  where $g_{\small\mathrm{hZZ}}$ is the non-standard hZZ coupling and
  $g^{\small\mathrm{SM}}_{\small\mathrm{hZZ}}$ is the same coupling in the SM. The
  Higgs boson is assumed to decay exactly as in the SM, but the
  production cross-sections for the Higgsstrahlung and fusion
  processes are scaled with $g_{\small\mathrm{hZZ}}^2$\,\cite{lep_smh}.}
\end{minipage}
\hspace{2pc}%
\begin{minipage}{18pc}%{14pc}
\includegraphics[width=15pc]{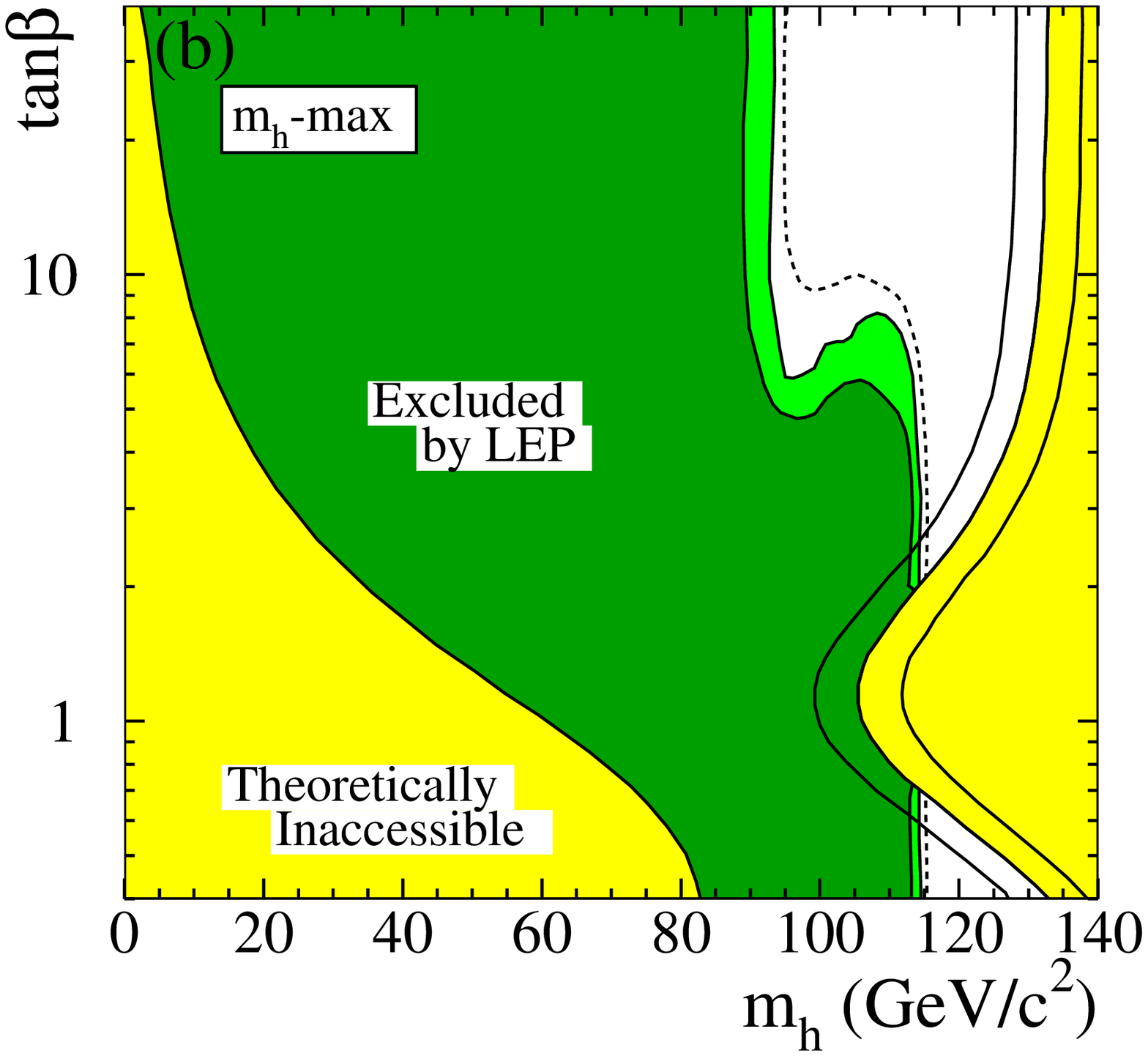} %18pc
\caption{\label{fig-two}Exclusions at the 95\%\,CL (light green) and
  99.7\%\,CL (dark green) in the $(\mh,\tan\beta)$
  plane\,\cite{lep_mssmh} for $\mt=174.3\,\Gcs$. The dashed line is
  the expected 95\%\,CL exclusion. Theoretically inaccessible regions
  are shown in yellow. The parameter space upper boundary on $\mh$ is
  drawn for (from left to right) $\mt=169.3,\,174.3,\,179.3\,\Gcs$.}
\end{minipage} 
\end{figure}
\end{center}

No single LEP experiment has the power to distinguish between the two
hypotheses at more than the two-sigma level, for an SM Higgs signal
with mass larger than $\sim$114\,$\Gcs$. Most of the discriminating
power is concentrated in the combined four-jet channels, whose
discriminating power roughly equals that of all the other channels put
together.
For a test mass $\mh=115\,\Gcs$ the highest weight candidate events
observed in the data\,\cite{lep_smh} are three four-jet candidates
recorded by ALEPH, and one missing energy candidate recorded by L3.
At this test mass, the ALEPH result\,\cite{a_smh} corresponds to a 3
sigma excess; the L3 result\,\cite{l_smh}, while consistent with the
background hypothesis, very slightly favours the
signal-plus-background hypothesis; the results of DELPHI\,\cite{d_smh}
and OPAL\,\cite{o_smh} are consistent with the background-only
hypothesis. The combined LEP result at $\mh=115\,\Gcs$ is in excess of
the background-only hypothesis by 1.7 standard deviations.
The lower bound on the SM Higgs boson mass at the 95\%\,CL determined
from the LEP-combined result is $\mh=114.4\,\Gcs$, while the median
expected limit is $115.3\,\Gcs$. SM Higgs boson signals of lower mass
are excluded at higher confidence levels. 95\%\,CL upper bounds on the
hZZ coupling in non-standard models were also obtained, from the LEP1
and LEP2 Higgs searches (Fig.\,\ref{fig-one}).
% ------
% ------
\mbox{ }\\[-1.75cm]
\section{The search for Higgs bosons beyond the SM}
The Higgs sector of the Minimal Supersymmetric extension to the SM
(MSSM) predicts 3 neutral ($\h,\,A,\,H$; $\mh<\mH$) and 2 charged
($\Hpm$) scalars. The lightest CP-even Higgs boson, whose mass is
predicted to be less than $\sim140\,\Gcs$, can be produced by
Higgsstrahlung; a complementary production process is pair production
in association with the CP-odd $\A$ ($\ee\r\h\A$). The final results
of the extensive searches for the neutral MSSM Higgs bosons at LEP are
reported in detail in Ref.\,\cite{lep_mssmh} for a number of benchmark
scenarios, including CP-conserving/CP-violating Higgs sectors. For
instance, in the CP-conserving $\mh-$max benchmark the following are
excluded at the 95\%\,CL or higher: $\mh<92.9\,\Gcs,\,\mA<93.4\,\Gcs,$
and $0.9\,(0.6)<\tan\beta<1.5\,(2.6)$ for $\mt=179.3\,(169.3)\,\Gcs$
(Fig. \ref{fig-two})\footnote{$\tan\beta$ is the ratio of the vacuum
  expectation values of the two Higgs doublets.}. In the case of
CP-violating scenarios the Higgs mass limits can be severely weakened,
especially for $\tan\beta\gsim{3}$.

The results of the searches for pair-produced charged Higgs bosons are
reported in\,\cite{lep_ch}: $\mHpm<78.6\,\Gcs$ is excluded by the data,
at the the 95\%\,CL.
Finally, searches for neutral Higgs particles with non-standard decay
modes were also performed, and strong 95\%\,CL lower bounds were set
on the Higgs mass; details can be found elsewhere: Higgs decaying
exclusively to $\bb$ or to $\tautau$\,\cite{lep_smh}, fermiophobic
Higgs\,\cite{lep_fermio}, invisible Higgs\,\cite{lep_inv} and
flavour-independent Higgs decays\,\cite{lep_flavind}.
\ack I would like to thank the EPS HEP2007 organisers for such an
interesting and well-organised conference, and also my colleague Gavin
Davies for assistance in preparing this talk.
\end{document}